\documentclass[aps,pra,twocolumn,superscriptaddress,longbibliography]{revtex4-1}
 \usepackage{graphicx}
\usepackage{latexsym}
\usepackage{amssymb}
\usepackage{amsmath}
\usepackage{amsfonts}
\usepackage{upgreek}
\usepackage{float}
\usepackage{bm}
\usepackage{multirow}
\usepackage{color}
\usepackage[T1]{fontenc}
\usepackage{hyperref}
\usepackage{subfigure}
\usepackage [latin1]{inputenc}

\hypersetup{
colorlinks = true,
linkcolor = [rgb]{0.70,0.13,0.13},
citecolor = [rgb]{0.13,0.55,0.13},
urlcolor  = [rgb]{0.25, 0.41, 0.88}}

\begin{document}

\title{Quantum criticality and universality in the $p$-wave paired Aubry-Andr\'{e}-Harper model}

\author{Ting Lv}
\affiliation{College of Science, Nanjing University of Aeronautics and Astronautics, Nanjing, 211106, China}
\affiliation{Key Laboratory of Aerospace Information Materials and Physics (NUAA), MIIT, Nanjing 211106, China}

\author{Tian-Cheng Yi}
\affiliation{Beijing Computational Science Research Center, Beijing 100193, China}

 \author{Liangsheng Li}
 \affiliation{Science and Technology on Electromagnetic Scattering Laboratory, Beijing 100854, China}

\author{Gaoyong Sun}
\affiliation{College of Science, Nanjing University of Aeronautics and Astronautics, Nanjing, 211106, China}
\affiliation{Key Laboratory of Aerospace Information Materials and Physics (NUAA), MIIT, Nanjing 211106, China}

\author{Wen-Long You} \email{wlyou@nuaa.edu.cn}
\affiliation{College of Science, Nanjing University of Aeronautics and Astronautics, Nanjing, 211106, China}
\affiliation{Key Laboratory of Aerospace Information Materials and Physics (NUAA), MIIT, Nanjing 211106, China}

\begin{abstract}
We investigate the quantum criticality and universality in Aubry-Andr\'{e}-Harper (AAH) model with $p$-wave superconducting pairing $\Delta$ in terms of the generalized fidelity susceptibility (GFS). We show that the higher-order GFS is more efficient in spotlighting the critical points than lower-order ones, and thus the enhanced sensitivity is propitious for extracting the associated universal information from the finite-size scaling in quasiperiodic systems. The GFS obeys power-law scaling for localization transitions and thus scaling properties of the GFS provide compelling values of critical exponents. Specifically, we demonstrate that the fixed modulation phase $\phi=\pi$ alleviates the odd-even effect of scaling functions across the Aubry-Andr\'{e} transition with $\Delta=0$, while the scaling functions
for odd and even numbers of system sizes with a finite $\Delta$ cannot coincide
irrespective of the value of $\phi$.
A thorough numerical analysis with odd number of system sizes reveals the correlation-length exponent $\nu$$\simeq$ 1.000 and the dynamical exponent $z$ $\simeq$ 1.388 for transitions from the critical phase to the localized phase,
suggesting the unusual universality class of localization transitions in the AAH model with a finite $p$-wave superconducting pairing lies in a different universality class from the Aubry-Andr\'{e} transition. The results may be testified in near term state-of-the-art
experimental settings.
\end{abstract}

\date{\today}
\maketitle

\section{Introduction}
\label{Sec1:Introduction}
Quantum phase transitions (QPTs) have attracted the intense interest of both theorists and experimentalists in condensed matter physics for decades.
With variation of a non-thermal variable in the many-body Hamiltonian,
the ground-state properties show abrupt changes as a result of competing ground-state phases ~\cite{RevModPhys.69.315,2003Quantum,Shen2020}.
The Landau-Ginzburg-Wilson (LGW) paradigm has provided a well-established framework,
which rests on the assumption that a continuous equilibrium phase transition can be accessed by the onset of an order parameter associated with spontaneous symmetry breaking. A broad regime of order parameter fluctuations are controlled by proximity to a quantum critical point (QCP).
It has been a long time to recognize that the symmetry breaking occurs spontaneously only in infinite-size quantum many-body systems, while recently they are sometimes observed in surprisingly small systems~\cite{Bayha_2020} and even few-body systems~\cite{Cai_2021,PhysRevLett.119.220601,PhysRevLett.115.180404}.
Despite the enormous success, the breakdown of the LGW paradigm appears in a few different situations, for example, deconfined QCPs~\cite{Wang_2017,Shao_2016} and deconfined critical universality classes~\cite{PhysRevResearch.2.023031}.

In recent years, an immense effort has been expended to understand QPTs in quasiperiodic systems~\cite{RevModPhys.65.213,Iyer_2013,PhysRevB.35.1020,2009PhT.62h.24L,2019Natur.577.42W,goncalves2021hidden,PhysRevLett.110.176403,2005ChPhL.22.2759G,Agrawal_2020,PhysRevLett.126.106803,Goblot_2020,PhysRevLett.125.196604}.
The quasiperiodic system possesses a long-range periodicity, which
is intermediate between that of the clean and randomly disordered cases,
offering a rich playground to study quasiperiodic QCPs and unusual characteristic features, such as hierarchical energy spectra~\cite{PhysRevB.34.5208,1992ELECTRONIC,PhysRevB.40.8225} and  localization-delocalization transitions. A paradigmatic model of the quasiperiodic system is the Aubry-Andr\'{e}-Harper (AAH) model~\cite{Liu_2017,Zeng_2018,PhysRevB.103.104202,PhysRevB.91.014108,2018PhRvB..97q4206P,PhysRevB.94.125408,Wei2019FidelitySI,PhysRevB.100.195143,PhysRevResearch.3.013148,xiao2021observation,PhysRevB.93.104504,PhysRevB.91.014108,PhysRevB.90.104204,PhysRevB.102.195142,PhysRevA.95.062118,PhysRevA.95.043837,PhysRevB.100.125157,PhysRevB.101.020201,PhysRevB.103.054203,PhysRevB.101.235150,PhysRevB.100.125157,PhysRevB.102.024205}, in which the quasiperiodicity is embodied in the form of a cosine modulation incommensurate with lattice spacing.
With the rapid development of experimental technologies, the AAH model can be realized in optical waveguide lattices~\cite{PhysRevLett.62.977,2003Natur.424..817C,2004PhRvL..93e3901P}, photonic crystals~\cite{PhysRevLett.90.055501,PhysRevLett.103.013901,PhysRevLett.109.106402}
and cold atom systems~\cite{Roati_2008}. For instance,
the AAH Hamiltonian has been experimentally realized by cold atomic gases in a one-dimensional optical lattice perturbed by another weak incommensurate optical lattice~\cite{Roati_2008}.
These feasible platforms allow us to explore the emerging topological states of matter
with additional interactions in incommensurate systems, including
modulated off-diagonal hopping~\cite{PhysRevB.91.014108},
nearest neighbor $p$-wave superconductivity~\cite{PhysRevB.100.064202}, a long-range $p$-wave superconducting pairing~\cite{PhysRevResearch.3.013148}, and many-body interactions~\cite{PhysRevB.102.195142,2021arXiv210613841S}.

The AAH model has gained popularity since it acts as a proxy for
random potentials in the study of generic disordered system.  An obstacle to comprehending the critical
phenomena in the disordered systems is the undecidability of local order parameters.
In close proximity to QCPs, the complex and non-local
entanglement between individual constituents becomes extremely prominent at all distance scales.
As such, it has been recognized that the exploration of quantum critical phenomena from the perspective of quantum information science is a great privilege, such as the von Neumann entropy~\cite{PhysRevB.78.115114}, and quantum concurrence~\cite{PhysRevLett.105.095702}.
The quantum fidelity susceptibility (QFS) has proved to be particularly useful for detecting the critical points of a symmetry-knowledge unknown system~\cite{PhysRevA.97.013845,2009PhRvL.102e7205G,PhysRevA.77.032111,PhysRevB.103.014446}.  It was shown that the QFS can not only identify the QCPs, but also satisfy the scaling ansatz, where the universal information can be retrieved.
The most significant implication is that the finite-size scaling of such a universal order parameter dictates position of QCPs and  the critical exponent of the correlation
length $\nu$. For a quasi-periodic system with spatial complexity, the scaling theory of the QFS and the universality of localization transition have been partially understood.  Notably, the finite-size scaling of the usual fidelity susceptibility is irrelevant to the dynamical exponent $z$. However, the critical exponents obey the scaling and hyperscaling relations, implying that there are only two independent exponents. Thus, a second independent critical exponent plays a decisive role in determining the universality class, which lies at the heart of critical phenomena. In this work, we apply the generalization of fidelity susceptibility to the one-dimensional AAH model with $p$-wave superconducting pairing,  and devise a direct pathway to the determination of critical points and universal critical exponents of localization-delocalization transitions. Importantly, the theoretical predictions could be testified in state-of-the-art experiments.

The rest of the paper is organized as follows.  Section~\ref{sec:model} reviews the AAH model with $p$-wave superconductivity and determines its phase diagram. In Sec.~\ref{sec:Fidelity}, we introduce the concept of the generalized fidelity susceptibility (GFS) and postulate its scaling
hypothesis for the universal part. Section~\ref{sec:SUSCEPTIBILITY} is devoted to the scaling behavior of the GFS in the AAH model and identification of critical exponents. Conclusions and discussions are presented in Sec.~\ref{sec:SUMMARY}.

\section{MODEL HAMILTONIAN}\label{sec:model}
The generalized AAH model with $p$-wave superconducting pairing in a quasi-periodically modulated potential is given by the following Hamiltonian:
\begin{eqnarray}
H=\sum_{j=1}^N(-Jc_{j}^{\dag}c_{j+1}+\bigtriangleup c_{j}c_{j+1}+{\rm H.c})+
\sum_{j=1}^N V_{j}c_{j}^{\dag}c_{j},
\label{Ham}
\end{eqnarray}
where $c_{j}^{\dag}$ $(c_{j})$ is the fermionic creation (annihilation) operator at the $j$-th site among total $N$ lattice sites, $J$ is the hopping strength between nearest-neighbor sites, $\Delta$ denotes the amplitude of $p$-wave superconducting pairing, and H.c. represents the Hermitian conjugate. The $p$-wave pairing amplitudes can be tuned by the mixture of spin-polarized fermions with a Bose-Einstein condensate~\cite{PhysRevLett.121.253402}, affected by an $s$-wave Feshbach resonance in a spin-polarized cold Fermi gas~\cite{PhysRevA.94.031602}, or induced by the proximity effect in
stacking a superconducting wire
on top of the normal metal.
Here we
focus on quasi-periodicity encoded in the chemical potential, keeping a constant hopping magnitude and pairing potential.
The on-site potential terms are quasiperiodically varying  according to the Aubry-Andr\'{e} rule
$V_{j}=V\cos(2\pi\alpha j+\phi)$,
where $\alpha=(\sqrt{5}-1)/2$ is an irrational frequency and $V$ is the strength of the incommensurate potential. The parameter $\phi \in [0,2\pi)$ shifts the origin of the modulation representing a random phase. The boundary condition is imposed as $c_{N+1}$=$\sigma$$c_{1}$, where $\sigma$= $1$, $-1$, and $0$ corresponding to periodic, antiperiodic, and open boundary conditions, respectively. Without losing generality, $\Delta$ can be assumed to be
real [the phase can be otherwise eliminated under global U(1) transformation] and $J=1$ is set as energy unit throughout the paper.  For $\Delta=\pm 1$, the model will be equivalent to quasiperiodic Ising model~\cite{P2021Spin,PhysRevB.76.144427}.
When the $p$-wave pairing term is absent, i.e., $\Delta=0$, the AAH model in Eq.(\ref{Ham}) becomes easily tractable as it can be written as $H$ = $\sum_{i,j}$ $c_i^\dagger$ ${\cal H}_{i,j}$ $c_j$.
The eigenvectors $\vert \psi_n\rangle$ and the associated single-particle energies $\epsilon_n$ are obtained by diagonalizing the $N \times N$ single-particle Hamiltonian matrix ${\cal H}$.
In the limit when $V/J$ $\to$ 0, Eq. (\ref{Ham}) describes a metallic chain with all eigenstates being extended, while for $V/J$ $\to$ $\infty$ the eigenmodes are localized on one site. The Aubry-Andr\'{e} transition from being extended to being localized is known to occur at  $V/J=2$ as a consequence of Aubry-Andr\'{e} duality between the Hamiltonian in position and momentum space.

As for a finite $p$-wave paring, i.e., $\Delta$$\neq$$0$, the Hamiltonian (\ref{Ham}) can be diagonalized through a canonical Bogoliubov-de Gennes (BdG) transformation by introducing the new fermionic operators $\eta_{n}$ and $\eta_{n}^{\dag}$,
\begin{eqnarray}
\eta_{n}&=&
\sum_{j=1}^{N}(u_{n,j}^*c_{j}+v_{n,j} c_{j}^{\dag}),\quad
c_{j}=
\sum_{n=1}^{N}(u_{n,j}\eta_{n}+v_{n,j}^* \eta_{n}^{\dag}), \quad \quad
\end{eqnarray}
where $u_{n,j}$ and $v_{n,j}$ denote the two components of the wave function at site $j$, and $n$ ($n=1,...,N$) is the energy band index. The eigenstates $|\psi_n\rangle$= $(u_{n,1}$,$u_{n,2}$,...,$v_{n,1}$,$v_{n,2}$,...$)^{T}$
can be determined by solving the Schr\"{o}dinger equation $H |\psi_n\rangle=E_{n}|\psi_n\rangle$,
which can be recast into a $2N$$\times$$2N$ matrix form as
\begin{eqnarray}
 \left(
  \begin{array}{cc}
    A & B \\
    -B^* & -A^T
     \\
  \end{array}
\right)
\left(\begin{array}{c}
    u_{n,i}\\
    v_{n,i}^*
     \\
       \end{array}
     \right)= \epsilon_n \left(\begin{array}{c}
    u_{n,i}\\
    v_{n,i}^*
     \\
       \end{array}
     \right),
     \label{BdGHam}
\end{eqnarray}
where $A$ ($B$) is a $N\times N$ symmetric (antisymmetric) matrix. The nonzero elements
are given by $A_{i,i}$=$V_i$, $A_{i,i+1}$= $A_{i+1,i}$=$-J$, and $B_{i,i+1}$=-$B_{i+1,i}$=$\Delta$. The matrix elements for the boundary terms are $A_{N,1}$=$A_{1,N}$=$-\sigma J$, and $B_{N,1}$=-$B_{1,N}$=$-\sigma \Delta$. The BdG Hamiltonian ${\cal H}$ in Eq.(\ref{BdGHam}) respects an imposed  particle-hole symmetry, namely, $\tau^x {\cal H}^T\tau^x =-{\cal H}$,
where the Pauli matrix $\tau^x$ acts in the Nambu space. The energy levels appear in $\pm \epsilon_n$ conjugate pairs, with $\epsilon_{n}\ge0$, except the zero energy mode, which is self-conjugate.
As such, for finite lattices it is convenient to
replace $\alpha$ with
$F_{\ell-1}/F_{\ell}$,
the ratio of two successive Fibonacci numbers~\cite{1986PhRvB..34.7367M,PhysRevLett.51.1198}.
Note that the irrational limit is reached as far as the numerical results are extrapolated to the scaling limit
($\ell \to \infty$).
The period
$F_\ell$ then acts like
a finite length scale which controls scaling behavior. The Fibonacci-sequence quasiperiodic potential has an intimate connection with topological phase transition and Majorana modes.

\begin{figure}[tb]
  \centering
    \includegraphics[width=\columnwidth]{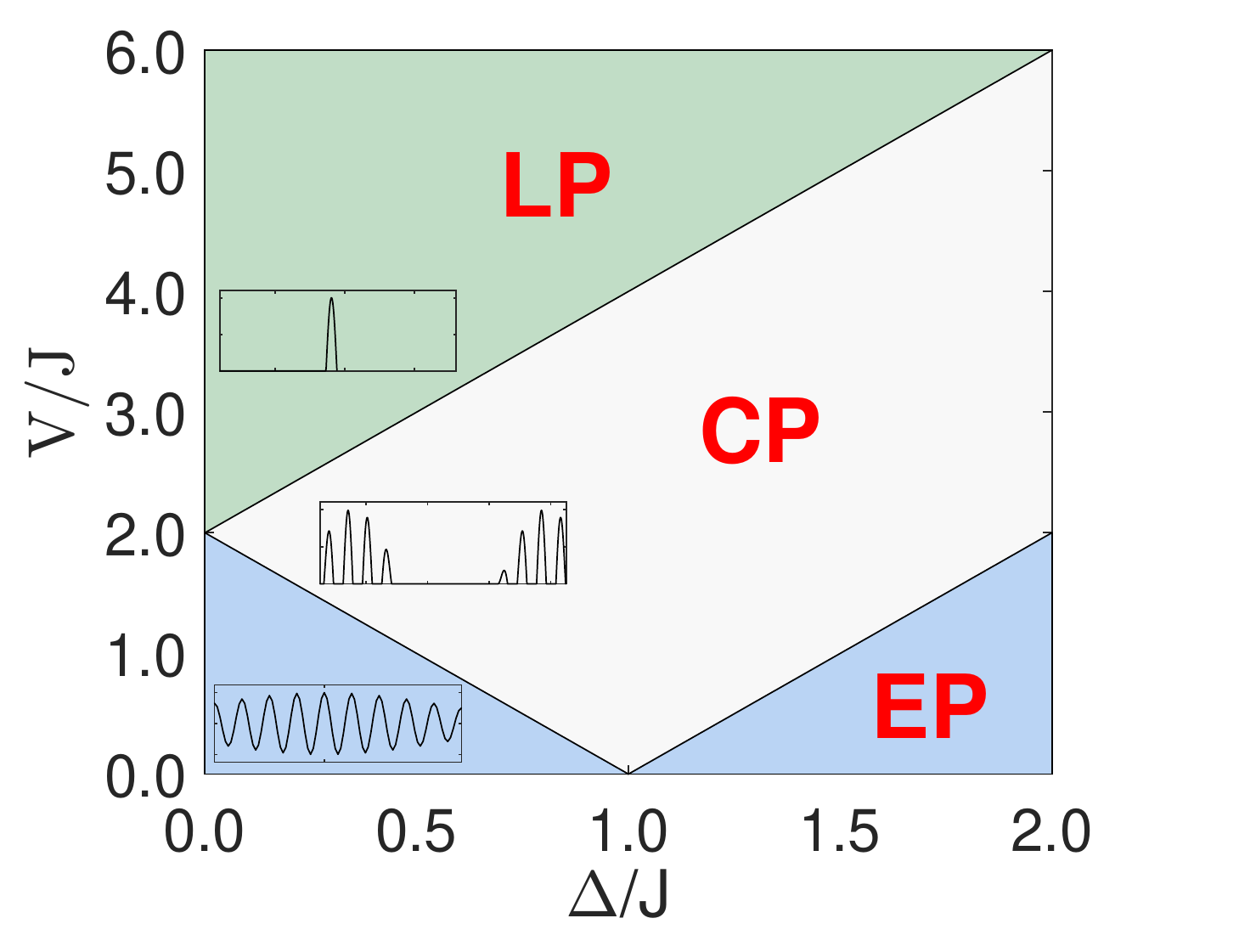}
 \caption{Phase diagram of the AAH model as a function of $p$-wave pairing strength $\Delta$ and incommensurate potential strength $V$. The localized phase (LP), critical phase (CP), and  extended phase (EP) are marked by blue, white, and green respectively. Three insets show the typical spatial distribution for localized, critical, and extended modes.  }
  \label{fig:phasediagram}
\end{figure}

While in the translational-invariant case the solution
of Eq.(\ref{BdGHam}) can be further reduced to the 2$\times$2 matrix form with independent momenta, in the quasidisorder
case one has to diagonalize the $2N$$\times$$2N$ BdG matrix
numerically, marking a qualitative difference between the
disordered and the clean model. In terms of the new fermion operators, the Hamiltonian in Eq.(\ref{Ham})
can be diagonalized as
\begin{eqnarray}
H=\sum_{n=1}^{N} \epsilon_{n} \eta^{\dag}_{n}\eta_{n} - \epsilon_{n}  \eta_{n}  \eta^{\dag}_{n} =\sum_{n=1}^{N}2\epsilon_{n}(\eta^{\dag}_{n}\eta_{n}-\frac{1}{2}),
\end{eqnarray}
with the single-particle eigenvalues being $\epsilon_{1}\le$$ \epsilon_{2}\le$ $\cdots$ $\le \epsilon_{N}$. The ground state of $H$ is the Bogoliubov vacuum state $|\psi_{g}\rangle$
annihilated by all $\eta_{n}$ for $n=1,...,N$, i.e.,  $\eta_{n} |\psi_{g}\rangle$=0, with an
energy $E_g$= -$\sum_{n=1}^{N} \epsilon_{n}$.
For a weak quasi-disorder strength, all the
eigenstates of the system are extended, while the system becomes localized for a sufficiently strong disorder. Recently, it was found that with a nonzero superconducting pairing, a nonergodic critical phase intervenes the transition from the delocalized to localized state, and
all the eigenstates are expected
to be multifractal.
 The phase diagram shown in Fig.\ref{fig:phasediagram} consists of the extended phase (EP), critical phase (CP), and localized phase (LP).  The localized wavefunctions in the LP can be transformed
into the extended ones in the EP by a Aubry-Andr\'{e} duality occurring at $\Delta = 0$ across a second-order QCP. The system undergoes a continuous QPT from the EP to the CP at
$V_{c1}=2|J-\Delta|$.  One finds that Eq.(\ref{Ham}) is invariant under
the transformation as $c_j$ $\to$ $-c_j^\dagger$ on odd $j$-th sites and $\alpha$ $\to$ $\alpha+1/2$~\cite{PhysRevB.93.104504}.
For the self-duality point ($J=\Delta$, $V=0$),
the EP becomes unstable for arbitrarily weak disorder. The system displays a second-order QPT from CP to LP at $V_{c2}=2|J+\Delta|$.

\begin{figure}[tb]
  \centering
  \includegraphics[width=\columnwidth]{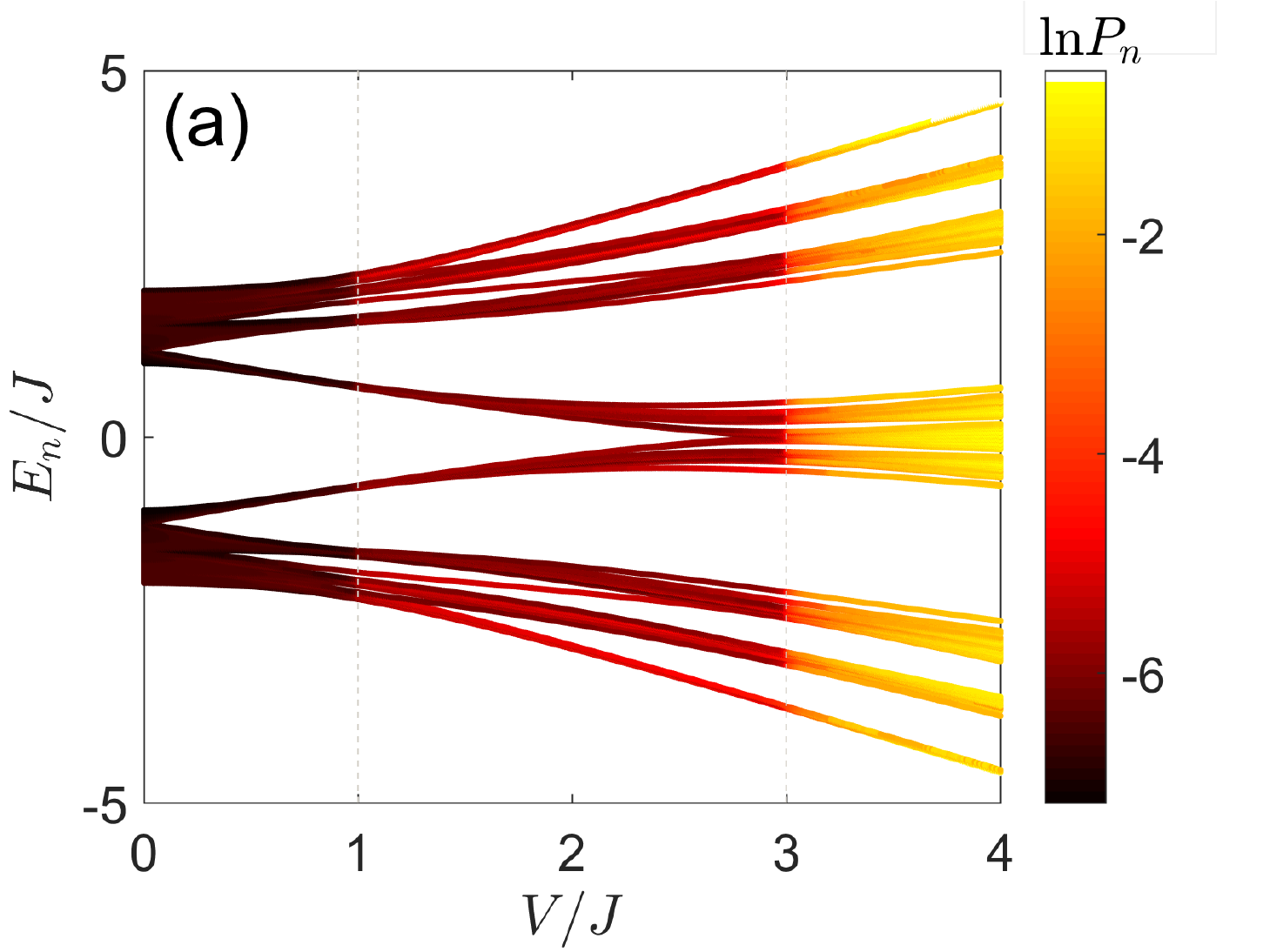}
  \includegraphics[width=\columnwidth]{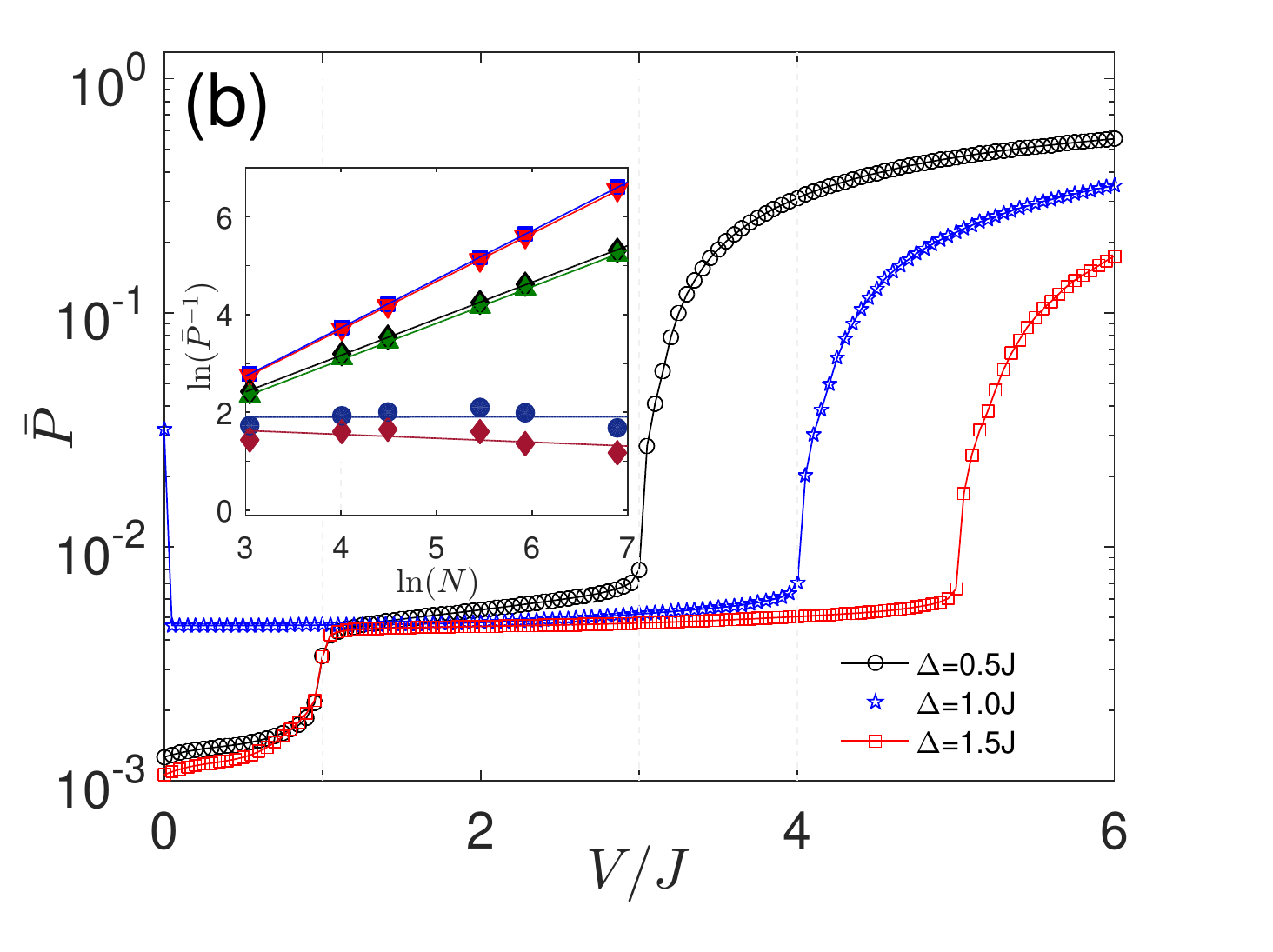}
 \caption{(a) Normalized IPR on a logarithmic scale for all eigenmodes of Eq. (\ref{Ham}) as a function of eigenenergies and $V/J$  with $\Delta$ = 0.5, $N=987$. The dashed lines mark the critical points at $V_{c1}$=$2|J-\Delta|$ and $V_{c2}$=$2|J+\Delta|$, respectively. The logarithmic scale is shown to have a better resolution. (b) MIPR as a function of the quasi-disorder strength at $\Delta=0.5J$ (black circles),  $\Delta=J$ (blue pentagons), and $\Delta=1.5J$  (red squares).
 Here we use the phase $\phi=\pi$ and the lattice site $N=987$. Inset shows the finite-size scaling of $\bar{P}^{-1}$ for $V=0.2J$, $0.5J$, $2.0J$, $2.5J$, $3.5J$, and $4.0J$ for $\Delta=0.5J$. }
 \label{fig:IPR}
\end{figure}

\begin{figure}[tb]
\centering
\includegraphics[width=\columnwidth]{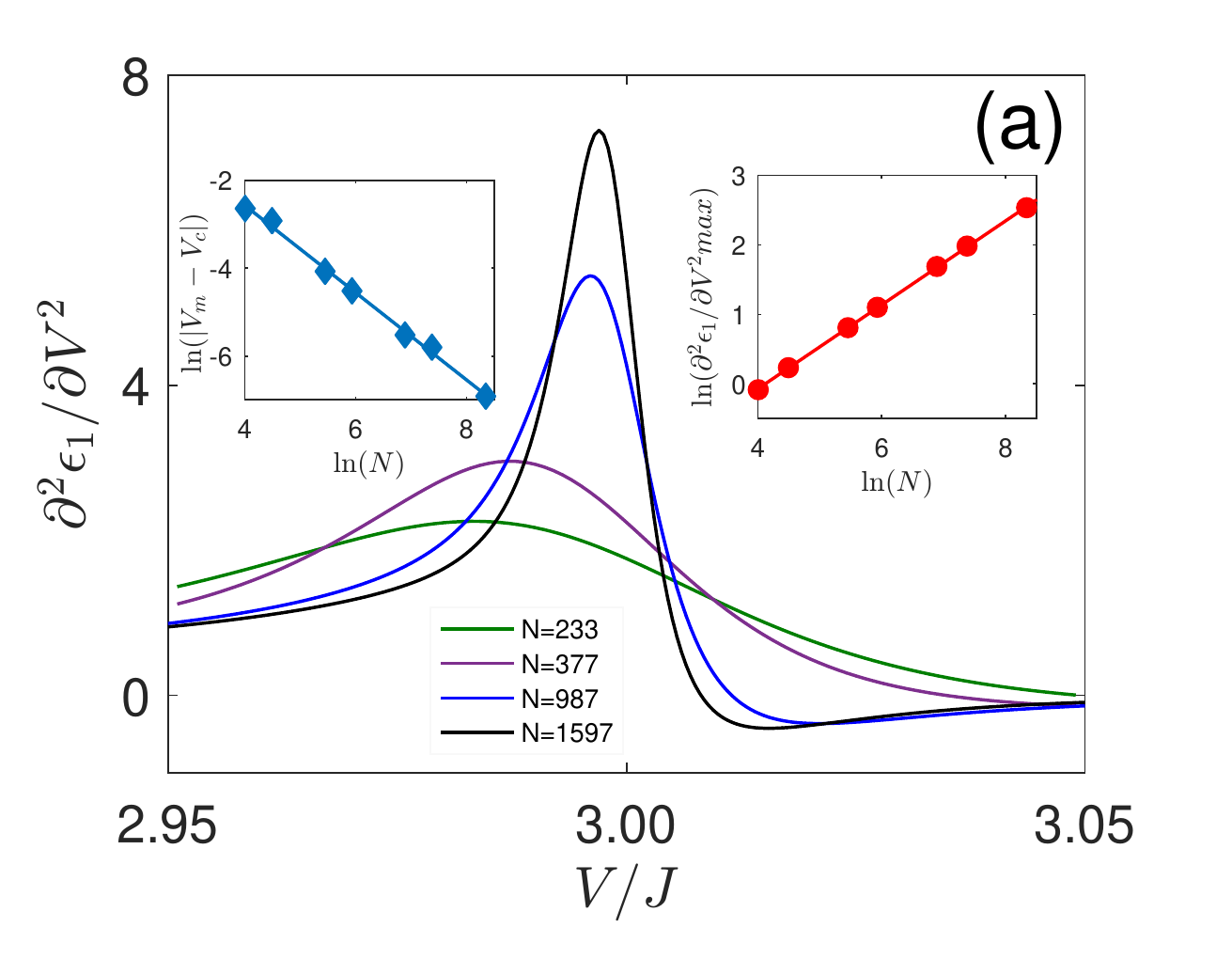}
\includegraphics[width=\columnwidth]{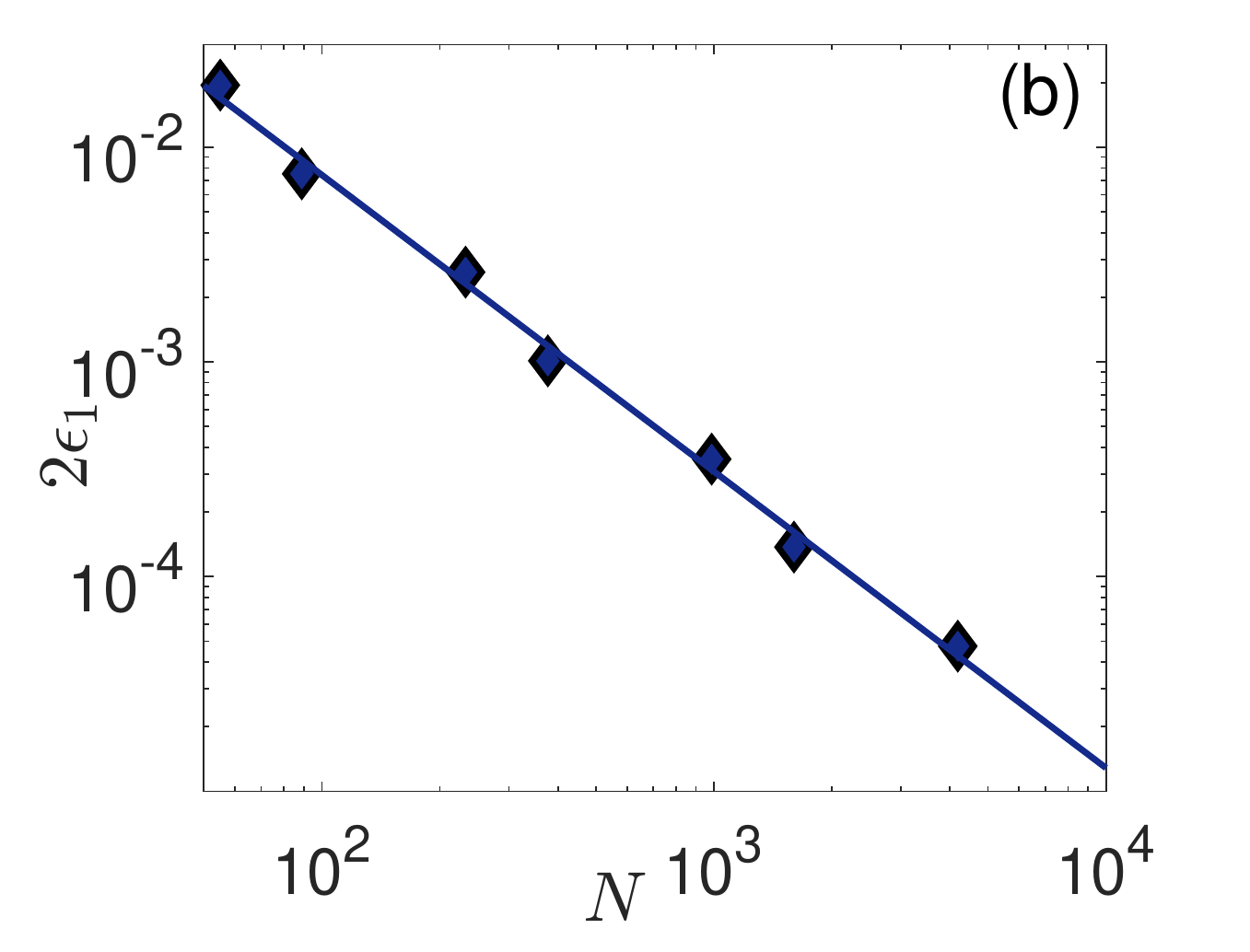}
 \caption{(a) The second-order derivative of first excitation $\epsilon_1$ with respect to $V$. Inset shows the scaling in the vicinity of the critical point $\ln (\vert  V_m-V_c\vert)$= $-0.994$($\pm 0.066$) $\ln N$ +1.408 ($\pm 0.410$)  and  the scaling behavior between the maximum value of $\ln (\partial^2 \epsilon_1/\partial V^2)_{\rm max}$ = 0.605 ($\pm 0.010 $) $\ln N$ $-2.491$($\pm 0.063$).   The symbols in the insets denote the numerical results and the solid lines correspond to the linear fittings. The parameters are $\Delta$=0.5, $\phi=\pi$.  (b) The excitation gap  $2\epsilon_1$ around the critical point $V_{c2}=3$ as a function of $N$ with $\Delta$=0.5, $\phi=\pi$.  }
\label{fig:epsilon1_Delta05_V3}
\end{figure}
In order to visually characterize the localized and extended nature of the entire energy spectrum, we evaluate the normalized inverse participation ratio (IPR) for each eigenstate $|\psi_{n}\rangle$ of the model~\cite{RevModPhys.80.1355,1978JPhC...11..925L,1980ZPhyB..36..209W,Misguich_2016}, given by
\begin{eqnarray}
P_{n}=\frac{\sum_{j}(u_{n,j}^{4}+v_{n,j}^{4})}{\sum_{j}(u_{n,j}^{2}+v_{n,j}^{2})}.
\end{eqnarray}
The IPRs can quantify the extent of distribution over the preferential bases.  It should be noted
that IPR is erroneously employed to describe the participation ratio in many literatures reports \cite{2005ChPhL.22.2759G,2009}, which is the reciprocal of IPR~\cite{Thouless197493}.
In a specific ${\cal N}$-dimensional bases $|\varphi_k\rangle$, the IPR of the state $|\psi_n\rangle=\sum_{k=1}^{\cal N} c_k|\varphi_k\rangle$ reaches a maximal value $P_\text{max}=1$
 when the state coincides exactly a single basis state, and attains a minimal value $P_{\rm min}=1/{\cal N}$
 when the state is uniform in the selective bases.  For a set of one-particle states in real space, the IPR scales inversely with the system size $N$ in the delocalized state, while appears to be independent of $N$ in the localized phase and shows intermediate behavior in the CP.
Both transitions at finite strength of quasiperiodic modulation occur simultaneously for
all eigenstates, as is revealed in Fig.\ref{fig:IPR}(a),
in contrast to the presence of mobility edges in specific systems~\cite{PhysRevB.83.075105,PhysRevLett.114.146601,PhysRevLett.120.160404,Guo_2020}. We thus further define the mean inverse participation ratio (MIPR) as
\begin{eqnarray}
  \label{eq:MIPR}
\bar{P}=\sum_{n=1}^{2N}
\frac{1}{2N}P_{n}.
\end{eqnarray}
The evolution of MIPR is exhibited in Fig.\ref{fig:IPR}(b) on a logarithmic scale for different $p$-wave
pairing strength $\Delta=0.5J$, $\Delta=1.0J$, and $\Delta=1.5J$.
One finds the MIPR is capable of identifying the phase boundaries separating the extended, critical, and localized phases, which are captured by the turning points of the MIPR locating respectively at $V_{c1}=2|J-\Delta|$ and $V_{c2}=2|J+\Delta|$. The LP is gapless for periodic boundary conditions while gapped for open boundary conditions.  Note that for a pure state $\rho=|\psi_n\rangle\langle \psi_n|$ of the entire system, the von Neumann entropy is zero and
the IPR is inversely proportional to the
participation entropy $S_q=\ln\sum_k\rho_{kk}^q/(1-q)$ of order $q=2$, i.e., $ S_2=-\ln P$, which becomes the diagonal entropy for the reduced density matrix of the subsystem~\cite{2011,PhysRevLett.107.040601}.
 As is observed in the inset of Fig.\ref{fig:IPR}(b), the scaling exponents extracted from the linear fittings for $V=0.2J$ and $V=0.5J$ with $\Delta=0.5J$ are approximately 0.98, which implies that the inverse of MIPR tends to scale extensively for extended states in the AAH model, resembling the volume-law scaling of the mean first-order R\'{e}nyi entropy at infinite temperature, which is conjectured to be universal for translationally
invariant
 quadratic fermionic Hamiltonians~\cite{PhysRevLett.121.220602},
while $\bar{P}^{-1}$ declines towards a finite value close to ${\cal O}(1)$ for localized states, in analogy to the area law of the disordered averaged entanglement entropy~\cite{2014PhRvL.113o0404P}. For critical states,
the MIPR scales like $N^{-d^*}$~\cite{1993Localization}, where the fractal dimension $0 < d^* < 1$ depends
on the fractal
structure of wavefunctions. The fitting lines of $\ln (\bar{P}^{-1})$ with respect to $\ln N$ for $V=1.5J$ and $V=2.0J$ with $\Delta=0.5J$ gives rise to $d^*$ $\approx$ 0.75, implying that
points in the whole CP belong to the same universality class.  Remarkably, the fractal dimension $d^*$ and the associated IPR are proved to host intrinsic relation to the mean entanglement entropy~\cite{PhysRevLett.124.200602}. One should be aware that a typical value of IPR can be used $\bar{P}_{\rm typ}$ = $ \sum_{n=1}^{2N} \ln P_{n}/(2N)$, which is similar to the behavior of Eq.({\ref{eq:MIPR}) for the AAH model (\ref{Ham}) yet becomes more subtle when eigenstates display a single-particle mobility edge.

\section{Generalized fidelity susceptibility and scaling hypothesis}\label{sec:Fidelity}

It is now well established that the QFS is a good measure to witness QPTs and manifest critical phenomena in translational invariant quantum systems~\cite{PhysRevA.78.012304,PhysRevE.74.031123,PhysRevE.76.022101}. The merit of the QFS in characterizing critical phenomena
is the model-independent feature, which is quite suitable for quantum systems without prior knowledge of order parameters. To this end, the fidelity susceptibility is recognized as a sensitive probe of quantum criticalities in conjugate field~\cite{PhysRevB.84.224435}, long-range interacting systems~\cite{hysRevA.98.023607,PhysRevB.101.094410}, deconfined QCP~\cite{PhysRevB.100.064427}, disordered systems,
 chaotic Hamiltonians~\cite{PhysRevX.10.041017}, quantum many-body scars~\cite{Surace_2021}, excited-state quantum phase transition~\cite{leblond2020universality}, and holographic models~\cite{PhysRevD.96.086004}.
Currently the investigations of the scaling of QFS in the context of quantum disordered systems are still poorly understood.  In what follows, we will  focus  on the fidelity susceptibility and its generalization as well as the associated scaling in the AAH model.

The fidelity susceptibility provides a generic and direct approach  to measure the quantum metric tensor via
the transition probability of the quantum state being
excited to other eigenstates during a sudden infinitesimal quench of
the tuning parameter $\lambda$~\cite{PhysRevE.76.022101}.
The GFS of order $2r+2$ at the tuning parameter $\lambda$ associated with the state $\vert\psi_n (\lambda)\rangle$ is given by~\cite{You_2015}
 \begin{eqnarray}
 \chi_{2r+2}^{(n)}(\lambda)= \sum_{m\neq n} \frac{\vert \langle  \psi_m (\lambda)  \vert \partial_\lambda \hat{H}  \vert \psi_n (\lambda) \rangle \vert^2}{[E_m(\lambda) -E_n(\lambda)]^{2r+2}},
\label{generalizedsusceptibility}
\end{eqnarray}
where $\vert\psi_m (\lambda)\rangle $ and $E_m (\lambda)$  correspond to the $m$th eigenstate and eigenvalue of this generic Hamiltonian $\hat{H}(\lambda)$,  respectively.
The numerator in Eq.(\ref{generalizedsusceptibility}) denotes the probability of exciting the system away
from the state $\vert \psi_n (\lambda)\rangle$ through a relevant (or marginal) perturbation $\partial_\lambda \hat{H}$. The GFS of different  orders is embodied  by  the  power  of  the denominator.
Concretely, Eq.(\ref{generalizedsusceptibility}) reduces respectively to the second derivative of the ground-state energy $\chi_1$ for  $r=-1/2$~\cite{PhysRevA.77.032111}  and the conventional quantum geometric tensor $\chi_2$ for $r=0$~\cite{PhysRevE.76.022101}. We can anticipate that $\chi_1$ has a weaker
divergence than $\chi_2$ at a critical point.
The QFS can be also devised as the Riemannian metric tensor upon projecting the dynamics onto a single (non-degenerate) band~\cite{PhysRevLett.99.100603},
\begin{eqnarray}
\chi_{2}^{(n)}(\lambda)=\langle \partial_\lambda \psi_n \vert  (1- \vert \psi_n\rangle \langle \psi_n \vert ) \vert \partial_\lambda \psi_n \rangle.\label{chipartialform}
\end{eqnarray}
Regarding the absence of mobility edge in the energy spectrum,
in the following we focus on the GFS of the lowest eigenstate $\vert\psi_1 (\lambda)\rangle $. To this end, the superscript of  $\chi_{2r+2}^{(1)}$ ($r$=$-1/2$, $0$, $1$) is omitted for abbreviation. 

We then start the description of the finite-size scaling theory by recalling its main features that hold in the vicinity
of the usual continuous QPT. The sensitivity is greatly enhanced, especially for the system at the quantum criticality compared with that away from the critical region~\cite{RevModPhys.90.035006}. Single parameter scaling posits that the correlation length $\xi$ is the only
relevant length scale in the thermodynamic limit that diverges
at the transition,
 \begin{eqnarray}
 \xi  \sim  \vert \lambda - \lambda_c\vert^{-\nu},
  \end{eqnarray}
and the single-particle spectral gap $\epsilon_s$ of size $2\epsilon_1$ will vanish as
 \begin{eqnarray}
 \label{Eq:Delta}
\epsilon_s \sim N^{-z},
\end{eqnarray}
 where $\nu$ is the correlation
length exponent and $z$ is the dynamical critical exponent. For finite chains, the
single parameter scaling hypothesis implies that the relevant
physical quantities shall depend only of the ratio
$\xi/N$, at least in the vicinity of the critical point where
$\xi \gg$ 1~\cite{PhysRevLett.28.1516}.
As $\lambda$ crosses QCPs adiabatically, the GFS shows a broad peak for a finite system size, signaling the location of pseudocritical points $\lambda_m$.
With increasing system sizes $N$, the peaks of GFS become more pronounced and the maximal points of the GFS is expected that
\begin{eqnarray}
\chi_{2r+2}(\lambda_{m})&\sim& N^\mu,
\label{eq:chi_max}
\end{eqnarray}
where $\mu=2/\nu+2zr$ is the critical adiabatic
dimension. For relevant operators $\partial_\lambda \hat{H}$ on sufficiently one-dimensional large-size systems, i.e., $\nu$$<$2, the pseudocritical points converge towards the critical points $\lambda_c$, satisfying
\begin{eqnarray}
\label{eq31} |\lambda_m-\lambda_c|\propto N^{-\theta},
\end{eqnarray}
with $\theta$=$1/\nu$. Here one should be aware that the shift exponent $\theta$ in Eq.(\ref{eq31}) is not necessarily equal to inverse of the
correlation-length exponent $\nu$~\cite{Roncaglia_2015}, as it happens to the entanglement witness~\cite{2002Natur.416..608O,Ren_2018}.
Accordingly, the GFS of a finite system with size $N$ in the neighborhood of a QCP shall obey the universal scaling form~\cite{PhysRevB.81.064418},
\begin{eqnarray}
\chi_{2r+2}(\lambda)=N^{2/\nu+2zr}\phi_{r}(\vert \lambda-\lambda_{m}\vert N^{1/\nu}),
\label{eq:chi_scaling}
\end{eqnarray}
where
$\phi_{r}(x)$ is a regular universal scaling function of the GFS of order $2r+2$, a priori unknown.
Estimates for critical parameters can thus be obtained by plotting
the scaled GFS $[\chi_{2r+2}(\lambda_m)$-$\chi_{2r+2}(\lambda)]$/$\chi_{2r+2}(\lambda)$
versus $N^{1/\nu}(\lambda-\lambda_{m})$  by subtly adjusting the values of $\lambda_{m}$, $\nu$, and $z$ until data collapse is achieved.
Alternatively, taking logarithm on both sides of Eqs.(\ref{eq:chi_max}) and (\ref{eq31}),yields
\begin{eqnarray}
\label{eq32} \ln |\lambda_m-\lambda_c| &\propto& c_{2r+2} \ln N, \nonumber\\
\ln  \chi_{2r+2} (\lambda_m)&\propto&d_{2r+2} \ln N,
\end{eqnarray}
where fitting parameters $c_{2r+2}$ and $d_{2r+2}$ can determine the critical exponents $\nu$ and $z$ as
\begin{eqnarray}
\label{eq33}
\nu&=& -1/c_{2r+2},  \quad
2 r z = d_{2r+2}+ 2 c_{2r+2}.
\end{eqnarray}

As such, we concentrate our attention on the region
close to the critical point of localization-delocalization transitions, above which the eigenstates are localized within the finite localization length $\xi$~\cite{PhysRevB.99.094203}.
Unlike the Aubry-Andr\'{e} model with $\Delta=0$, the BdG Hamiltonian (\ref{BdGHam}) for a finite $\Delta$ acts
in an enlarged expanded Nambu-spinor space~\cite{PhysRevB.55.1142}.  The second-order derivative of first excitation $\chi_1$$\equiv$$\partial^2 \epsilon_1/\partial V^2$ in the neighborhood of $V_{c2}$ for various odd number of system sizes $N$ with $\Delta=0.5$ is shown in Fig.\ref{fig:epsilon1_Delta05_V3}(a). One can see that $\chi_1$ presents a divergent peak ${\chi}_{1,\rm max}$ at $V_m$. The linear fittings give rise to parameters $c_{1}$ = -0.994 $\pm 0.066$  and $d_{1}$= 0.605  $\pm 0.010$. According to Eq.(\ref{eq32}), the fitting values imply that $\nu\approx1.006$ and  $z\approx1.384$.
Furthermore, the single-particle spectral gap $2 \epsilon_{1}$ with several system sizes from $N$ = 55 up to $N$ = 4181 are considered with periodic boundary conditions for $\Delta=0.5$ in Fig.\ref{fig:epsilon1_Delta05_V3}(b), indicating that $z =1.381$ $\pm 0.109$ according to Eq. (\ref{Eq:Delta}).

\section{FIDELITY SUSCEPTIBILITY IN
THE AUBRY-ANDR\'{E}-HAPER MODEL WITH \emph{P}-WAVE PAIRING}\label{sec:SUSCEPTIBILITY}
We obtain all the eigenenergies and the corresponding wave functions by
diagonalizing Eq.(\ref{BdGHam}). The numerical results tempt us to evaluate the GFS of the AAH model through Eq.(\ref{generalizedsusceptibility}).
First we recapitulate the finite-size scaling hypothesis of QFS in the Aubry-Andr\'{e} model with $\Delta=0$~\cite{PhysRevB.90.104204}.
Previous work revealed the QFS near the EP-LP transition can be separately rescaled onto two different universal curves for even and odd numbers of lattice sites~\cite{Wei2019FidelitySI}, while the retrieved critical exponents are quite close, which implies that two universal scaling functions are not necessary.
The logarithm of QFS $\ln \chi_{2} $ of the Aubry-Andr\'{e} model as a function of $V$ for $\Delta=0$ with different system sizes is exhibited in  Fig.\ref{fig:AAmodel}(a).
One can find that the fidelity
susceptibility presents a maximum $\chi_{2,{\rm max}} $ at $V_{m}$.
With increasing the system sizes, the peaks become more pronounced and $V_{m}$ gets closer to the critical point $V_{c}=2 J$. The maximum value of $\ln \chi_{2} $ against the system size $N$ is displayed in the log-log scale, whose linear fit shows that $\ln  \chi_{2,{\rm max}} $= $(2.000 \pm 0.010)$ $\times$$\ln N$-5.400, implying $\nu= 1.000 \pm 0.005 $ according to Eq.(\ref{eq:chi_max}). This is consistent with the Harris criterion~\cite{Harris_1974}, which imposes that $\nu<2$ for phase transitions
in the presence of incommensurate modulation.
When the rescaled fidelity susceptibility $[\chi_{2}(V_{m})$-$\chi_{2}(V)]$/$\chi_{2}(V)$ is plotted as a function of the proper scaling variable $N^{1/\nu}(V-V_{m})$,
all curves of distinct chain sizes in the vicinity of $V_{m}$
collapse into a single curve, as is shown in Fig.\ref{fig:AAmodel}(b),
which corroborates the estimated critical parameter
and the validity of the single-parameter scaling
hypothesis (\ref{eq:chi_scaling}). In particular the properly chosen value of $\phi=\pi$ much alleviates the odd-even effect.
\begin{figure}[tb]
  \centering
  \includegraphics[width=\columnwidth]{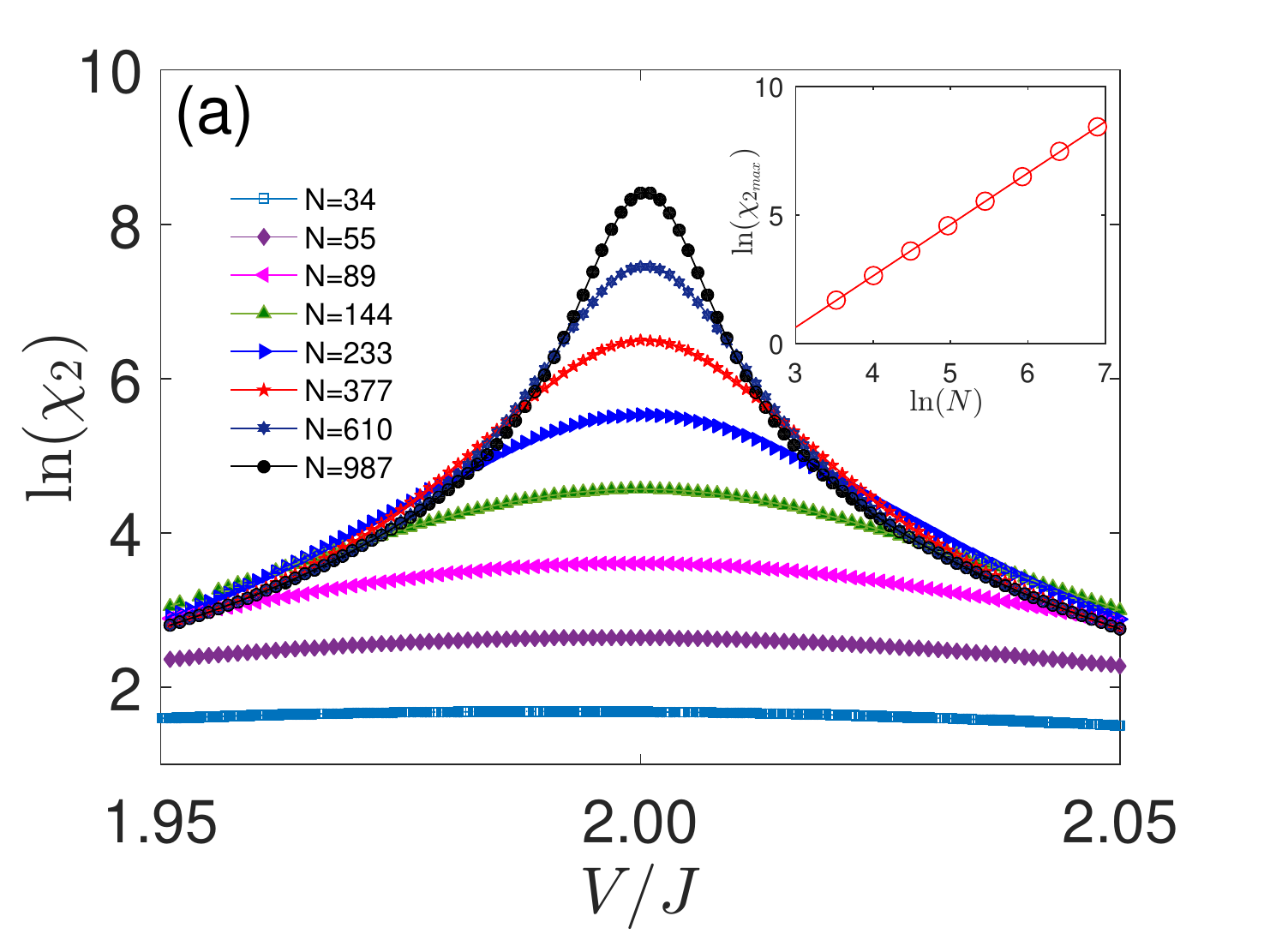}
  \includegraphics[width=\columnwidth]{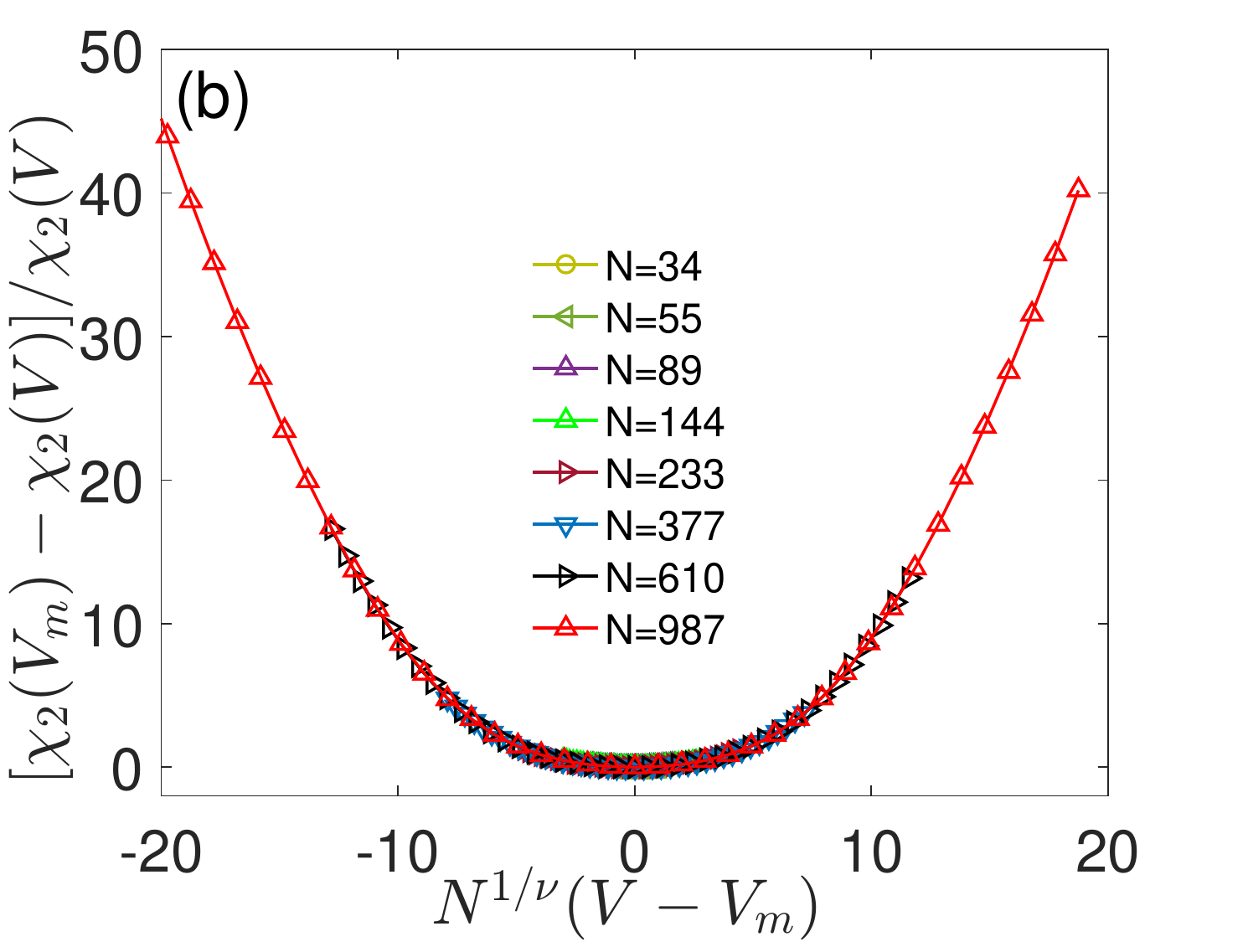}
\caption{ (a) The logarithm of the fidelity susceptibility $\chi_{2}$ as a function of the strength of the incommensurate potential $V$ for different number of lattice, and the logarithm of the maximum of fidelity susceptibility as a function of the logarithm of the system size, $N=34$, $55$, $89$, $144$, $233$, $377$, $610$, and $987$.  (b) Scaled fidelity susceptibility  $[\chi_{2}(V_m)-\chi_{2}(V)]/\chi_{2}(V)$  as a function of scaled variable $N^{1/\nu}(V-V_{c})$. All curves for number of the lattice sizes collapse into a single curve when we choose the correlation length critical exponents $\nu=1.000$. Here we choose we take periodic boundary conditions and set $\Delta=0$ and $\phi=\pi$.}
  \label{fig:AAmodel}
\end{figure}

Next, the QFS in the AAH model with respect to the strength of the incommensurate potential $V$ for odd number of lattice sizes with $\Delta=0.5$ is shown in Fig.\ref{fig:AAHmodeloddfs}. The QFS exhibits an extensive scaling in the off-critical region. Therefore, the QFS per site $\chi_2/N$ appears to be an $N$-independent value. Instead, the QFS shows a stronger dependence on system size around $V_{c2}=3$, signaling the onset of the QCP in the AAH model. The maximum values $ \chi_{2,\rm max}$ of the fidelity susceptibility near the QCP as a function of $N$ in log-log scale are plotted. The superextensive behavior at the pseudocritical point is reflected in the linear fitting $\ln \chi_{2,{\rm max}}$$\propto$(2.003 $\pm 0.033)$ $\ln N$,
whose slope suggests that $\nu$$=0.999\pm 0.017$. Meanwhile, the numerical fitting in terms of Eq.(\ref{eq31}) yields $\nu$=0.947 $\pm 0.090$. The accuracy of retrieved $\nu$ from the algebraic law (\ref{eq31}) is generally plagued by the precision of numerical calculation.

\begin{figure}[tb]
\centering
\includegraphics[width=\columnwidth]{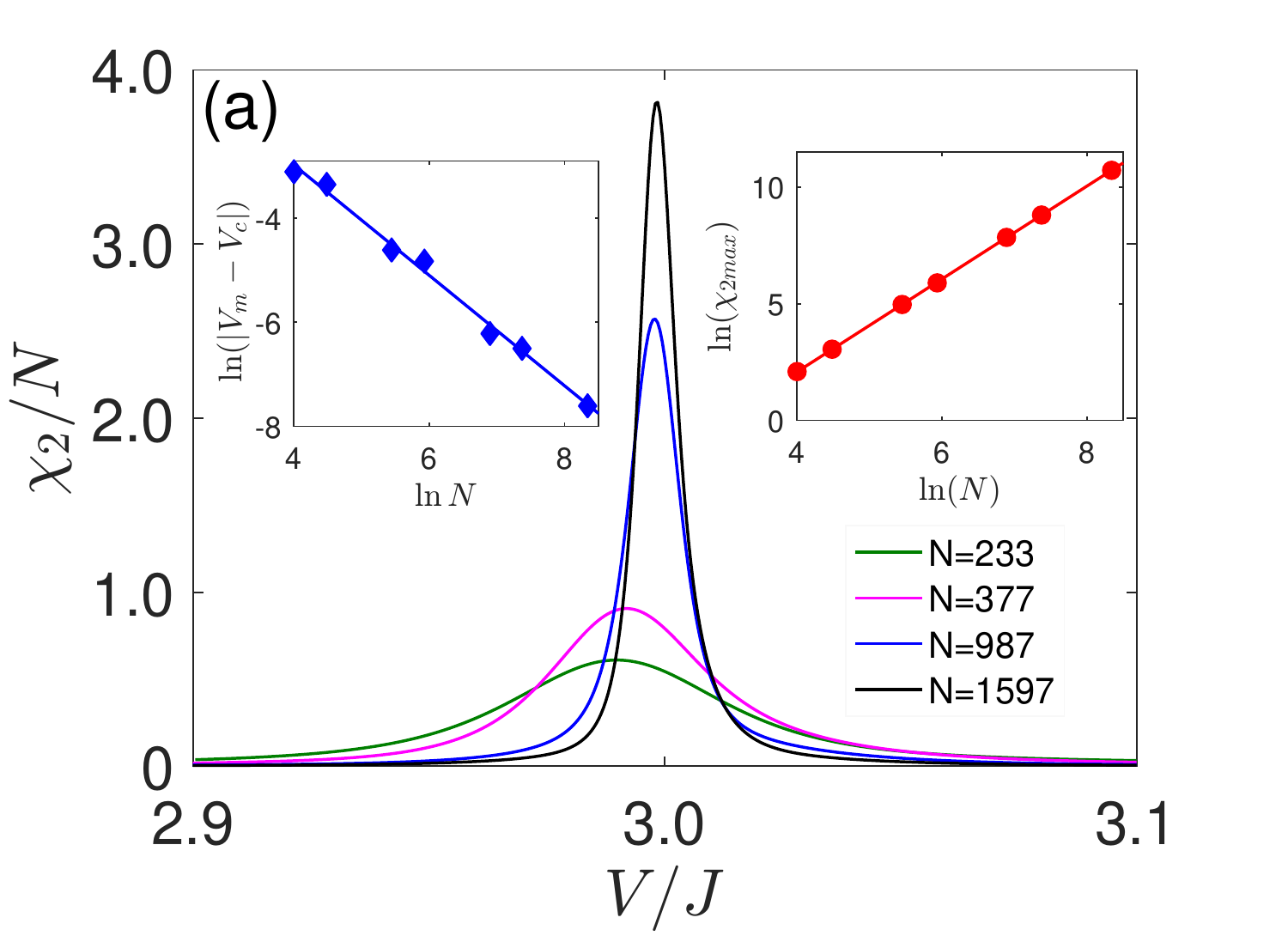}
\includegraphics[width=\columnwidth]{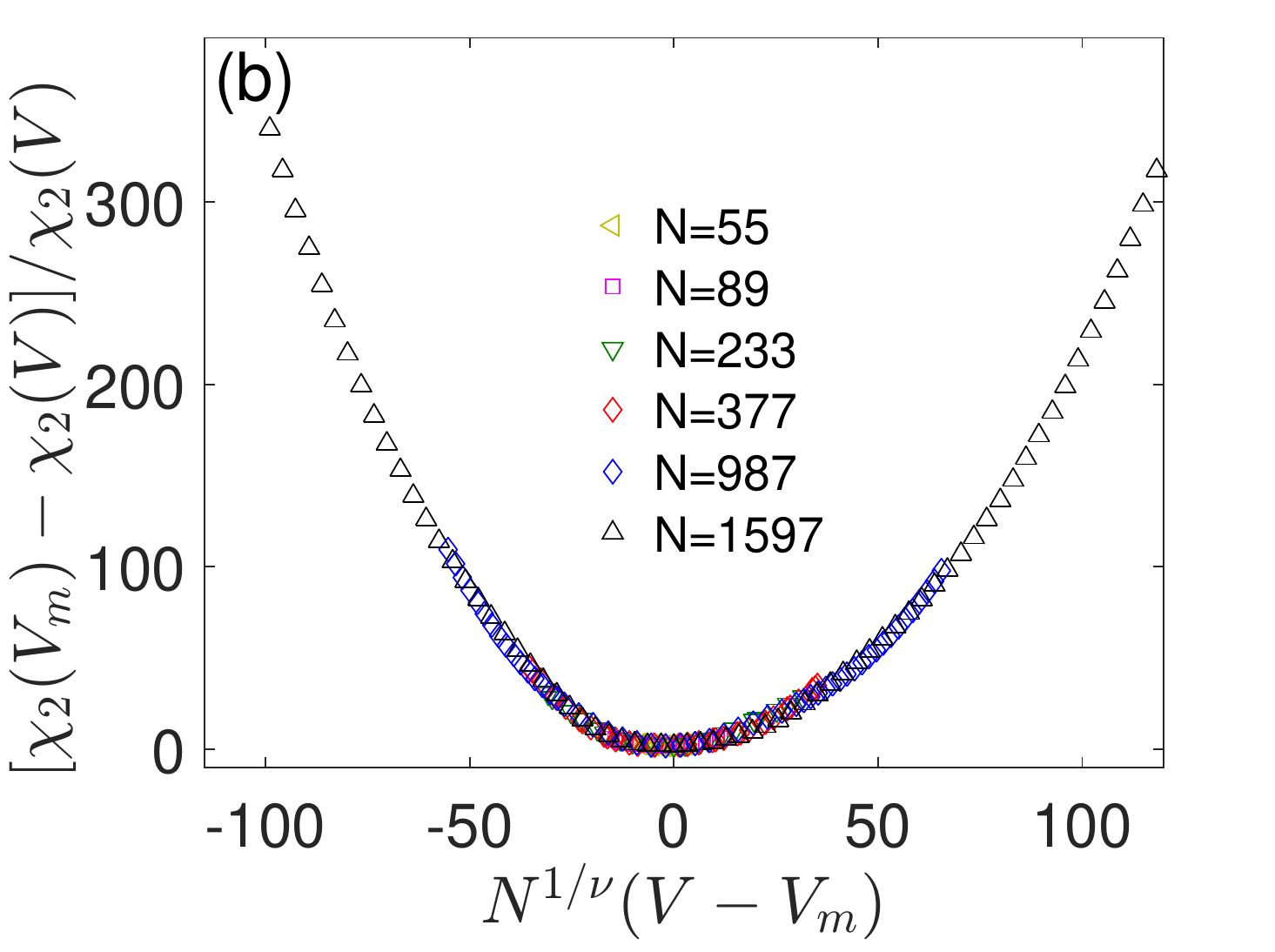}
 \caption{(a) The fidelity susceptibility per site $\chi_{2}/N$ as a function of the strength of the incommensurate potential $V$ with odd number of lattice sizes around $V_{c2}=3$. The inset shows the scaling behavior of the maxima versus the system sizes $N=55$, $89$, $233$, $377$, $987$, $1597$, $4181$. (b) Scaled fidelity susceptibility $[\chi_{2}(V_m)-\chi_{2}(V)]/\chi_{2}(V)$ as a function of scaled variable $N^{1/\nu}(V-V_{m})$. All curves for odd number of the lattice sizes collapse into a single curve when we choose the correlation length critical exponents $\nu=1.00$.
 Here periodic boundary conditions are used with $\Delta=0.5$ and $\phi=\pi$.}
  \label{fig:AAHmodeloddfs}
\end{figure}

In order to extract the dynamical exponent $z$ of CP-LP transitions, we further study the finite-size scaling of $\chi_{4}$. One can easily heed that $\chi_{4}$ displays much more divergent peaks than $\chi_2$ in the vicinity of QCP $V_{c2}=3$, as is disclosed in Fig.\ref{fig:AAHmodeloddgfs}.
The linear fittings of the peak maxima $\chi_{4,{\rm max}}$ suggests $c_4$=-1.018 $\pm 0.1074$ and $d_4=4.773 \pm 0.202$.  According to Eq.(\ref{eq33}) the extracted
values of critical exponents $\nu$ = $0.993\pm0.105$ and
$z = 1.380 \pm 0.053$ for the CP-LP transition with $\Delta=0.5$ agree well with those obtained from the gap scaling~\cite{PhysRevB.103.104202}. In this vein, we continue to pick the critical exponents $\nu$ and $z$ via the scaling analysis of $\chi_4$ as the $p$-wave superconducting pairing $\Delta$ changes. The numerical results in Fig. \ref{fig:AAHmodelcriticalexponents}(a) reveals that $\nu $$\simeq$ 1.000 and $z$ $\simeq$ 1.388 with little variation. It turns out that there is a discontinuity of $z$ when $\Delta$ increases from 0 to an infinitesimal value. Distinct values of $z \approx 2.375$ for $\Delta=0$~\cite{Wei2019FidelitySI} and $z \approx 1.380$ for $\Delta\ne 0$ suggest that their ground states belong to different universality classes.
For all $\Delta \neq 0$, the transitions across QCPs $V_{c2}=2|J+\Delta|$  belong to the same universality class as the quasiperiodic Ising chain~\cite{Agrawal_2020}.

\begin{figure}[tb]
  \centering
  \includegraphics[width=\columnwidth]{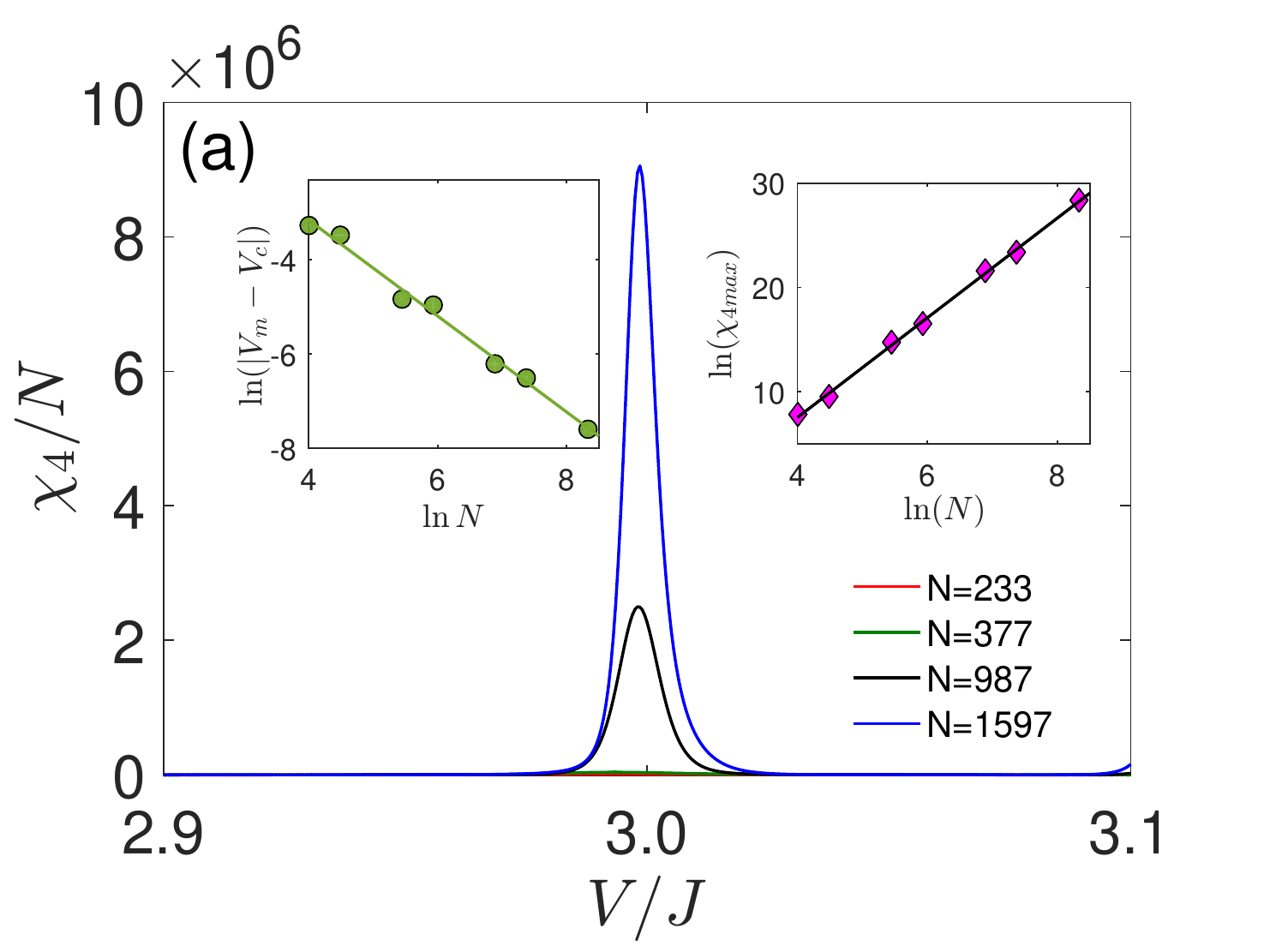}
  \includegraphics[width=\columnwidth]{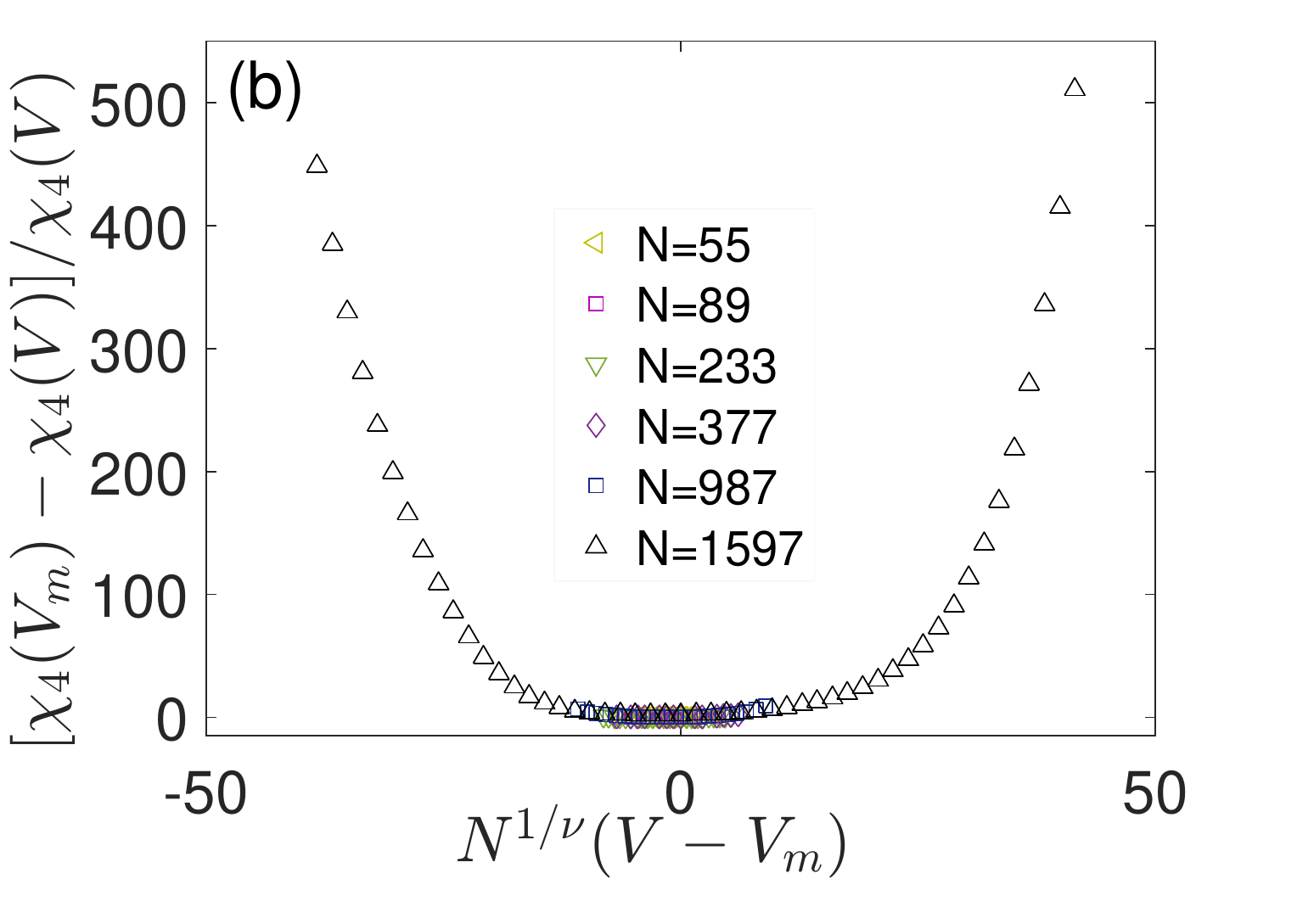}
   \caption{(a) The GFS per site $\chi_{4}/N$ as a function of the strength of the incommensurate potential $V$ with odd number of lattice sizes around $V_{c2}=3$. The inset shows the scaling behavior of the maxima versus the system sizes $N=55$, $89$, $233$, $377$, $987$, $1597$, and $4181$. (b) Scaled fidelity susceptibility  $[\chi_{4}(V_m)-\chi_{4}(V)]/\chi_{4}(V)$  as a function of scaled variable $N^{1/\nu}(V-V_{m})$. All curves for odd number of the lattice sizes collapse into a single curve when we choose the correlation-length critical exponents $\nu=1.00$. Here periodic boundary conditions are used with $\Delta=0.5$ and $\phi=\pi$. }
    \label{fig:AAHmodeloddgfs}
\end{figure}

\begin{figure}[tb]
  \centering
  \includegraphics[width=\columnwidth]{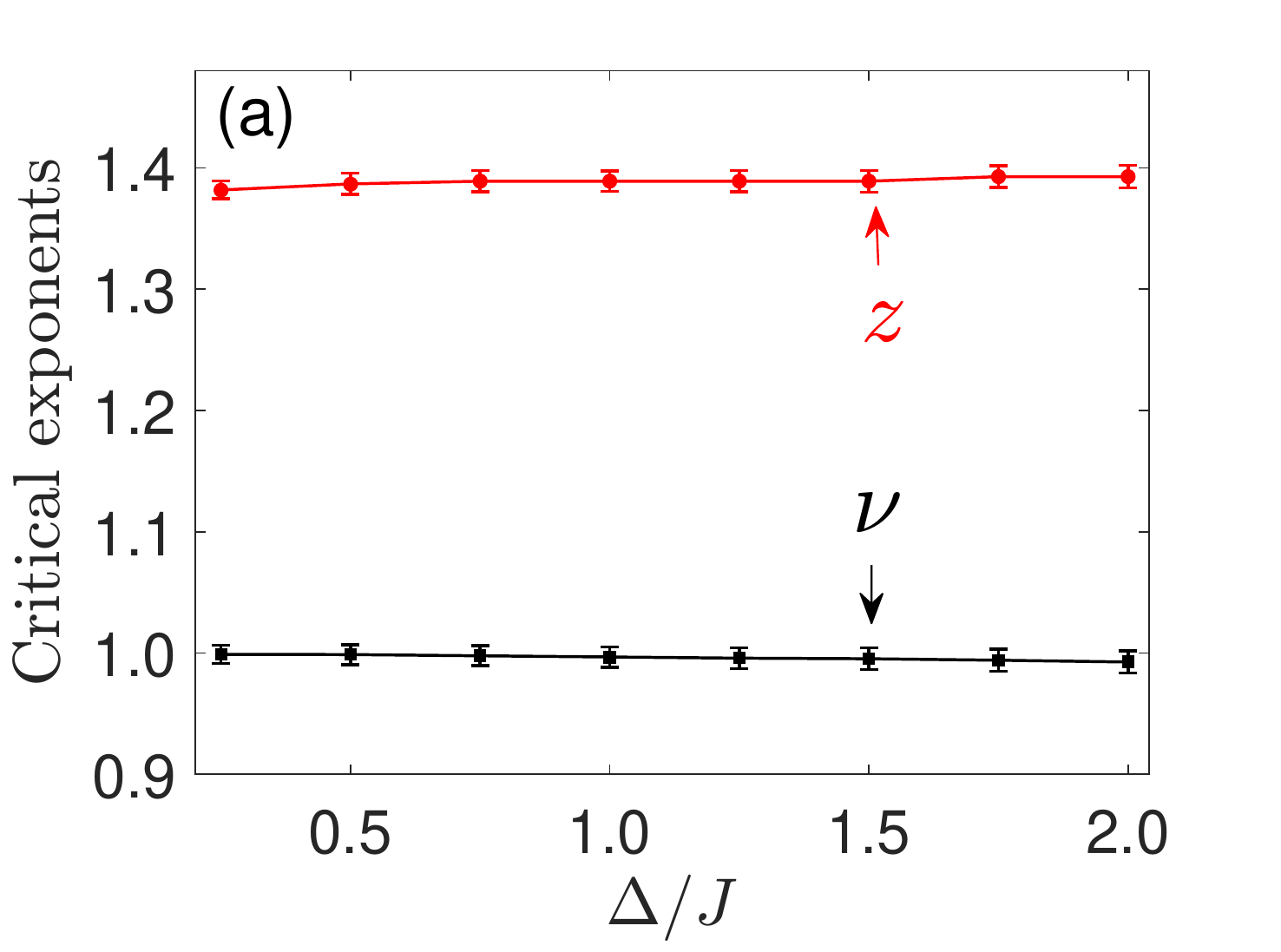}
  \includegraphics[width=\columnwidth]{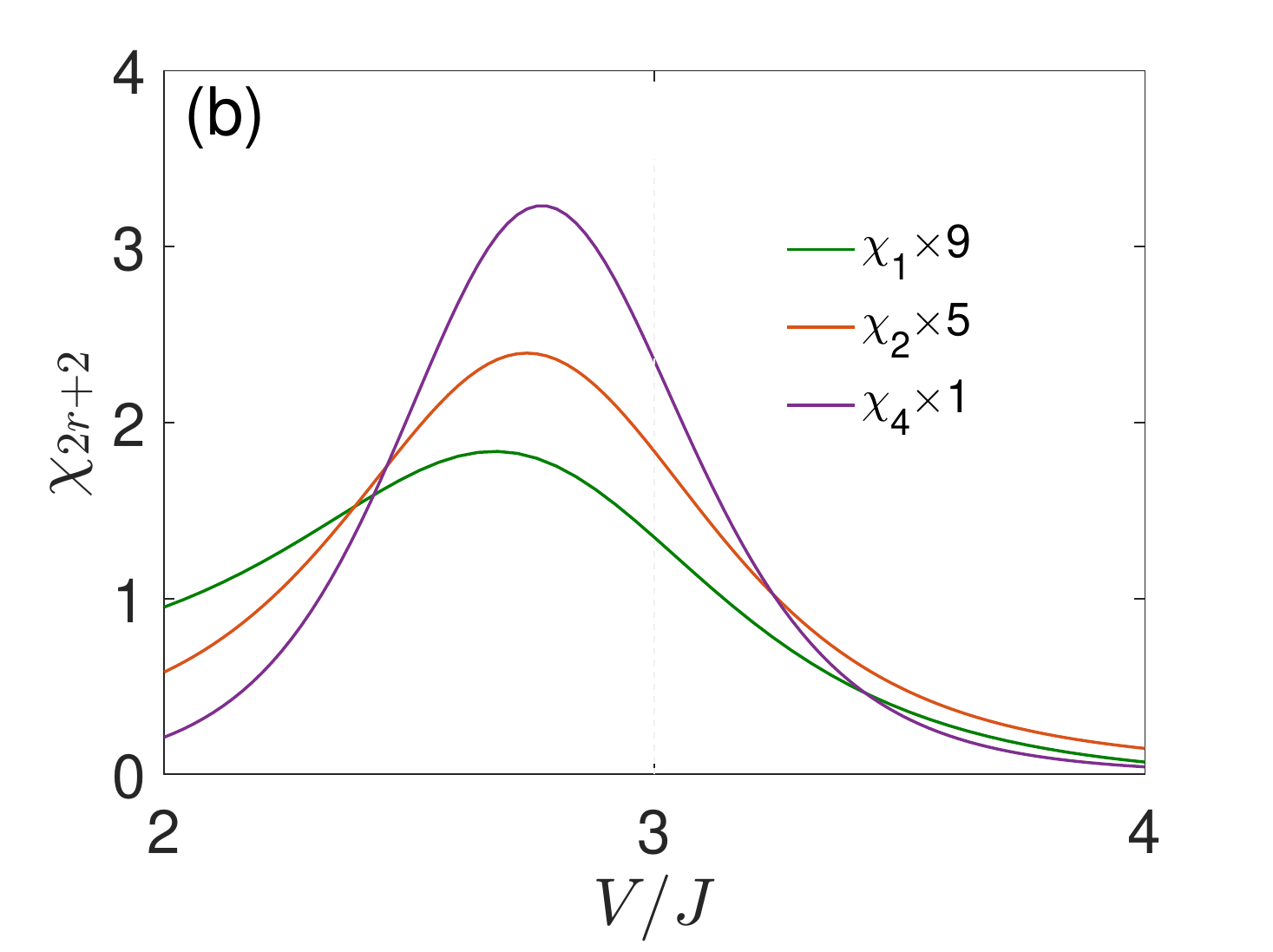}
   \caption{(a) The fitted values of critical exponents $\nu$ and $z$ as $\Delta$ varies. Here periodic boundary conditions are used with $\phi=\pi$. (b) The GFS $\chi_{2r+2}$ as a function of the strength of the incommensurate potential $V$  around $V_{c2}=3$ for a small system size $N=13$. Note that $\chi_1$ and $\chi_2$ have been respectively increased by a factor of 9 and 5 for guiding the eyes.}
    \label{fig:AAHmodelcriticalexponents}
\end{figure}

\section{Discussion and summary}\label{sec:SUMMARY}
In this work, we investigate quantum criticality in
the Aubry-Andr\'{e}-Harper (AAH) model with $p$-wave superconducting pairing in terms of the generalized fidelity susceptibility (GFS).
This quasiperiodic fermion system is Jordan-Wigner-equivalent to the quasi-periodically modulated transverse field\textit{ XY} chain. The interplay of spatial modulation of potential and symmetry breaking leads to quantum critical phenomena that are different from either the commensurate potential or randomly distributed potential.
In the absence of $p$-wave pairing ($\Delta$=0), the AAH model hosts a phase transition from the extended state to the exponentially localized state through the self-duality point ($V=2J$). With a finite value of $\Delta$, the transition
from the extended phase to the localized phase has to pass through an intermediate phase, and the critical point will develop into a critical region, which is sandwiched between the extended and localized states.
Various available methods have been incorporated in identifying quantum critical points (QCPs) from numerical simulations. A useful quantity in characterizing quantum criticality of disordered systems is the inverse participation ratio (IPR), which is equivalent to the second-order 
participation R\'{e}nyi entropy. Since there is no mobility edge in the energy spectrum, we then use the mean inverse participation ratio (MIPR) to characterize the degree of the extensivity in space of the wave function in different phases.
The MIPR presents a power-law scaling $\propto N^{-d^*}$ in distinct phases,
where $d^*$=1 in the extended phase, $d^*$=0 in the localized phase and the exponent $0<d^*<1$ in the intermediate critical phase.

We have developed accelerated methods for the location of critical points by the extrema of the universal order parameters. In this context, higher-order GFSs are more efficient in spotlighting the pseudo-critical points, even in the moderately large systems.
The enhanced sensitivity is propitious for extracting the associated universal information from the finite-size scaling in quasiperiodic QCPs, whose system sizes are rapidly
growing three-subsequence Fibonacci numbers~\cite{PhysRevB.101.174203}. This distinguishing feature becomes especially crucial in interacting many-body systems and  higher-dimensional systems.  One can see from Fig. \ref{fig:AAHmodelcriticalexponents}(b) that $\chi_4$ has already spied on the pseudocritical point for $N=13$ via the visible peak.  By performing a detailed numerical simulation, we find different orders of GFS obey power-law scaling in the vicinity of the localization transitions.
The single parameter scaling of these macroscopic observables provide self-consistent results of critical exponents. Moreover, the generalization of fidelity susceptibility poses an efficient avenue to dynamic exponent $z$.
The determined values of correlation-length exponent $\nu $$\simeq$ 1.000 and the dynamical exponent $z$ $\simeq$ 1.388 suggest that the quantum criticality of localization transitions in the AAH model for $\Delta \neq 0$ lies in a different universality class from the Aubry-Andr\'{e} transition ($\Delta=0$) with $\nu$=1.000, and $z$=2.375~\cite{Wei2019FidelitySI,PhysRevB.99.094203}, where a Aubry-Andr\'{e}-type duality may
prevent the finite energy excitations from localizing.  Understanding the nature of the quasiperiodic localization transition, with and without a finite $p$-wave superconducting pairing, may thereby cut to the heart of the phenomenon. The critical properties of this fixed point are found to be intermediate to the clean and randomly disordered
cases. The former case is represented by the clean transverse field Ising model in the celebrated Onsager universality class with $\nu$=$z$=1, while the latter is symbolized by the Anderson model with $\nu$=2/3, $z$=2~\cite{PhysRevA.84.055601}.

Last but not the least, another challenge in the study of quasiperiodic models is to separate
physically measurable observables from the mathematically
intriguing concepts. The quantum metric tensor has been experimentally measured with superconducting qubits~\cite{PhysRevLett.122.210401}, coupled qubits in diamond~\cite{Yu_2019}, and planar microcavity~\cite{Gianfrate2020}. Thus, an experimental measurement of the correlation-length exponent $\nu$ and the dynamical exponent $z$ becomes tractable. For instance, $z$ governs the low-temperature behavior of the specific
heat $C_v \sim T^{-z}$ and can be extracted from the density
of states $\rho \sim  \epsilon^{1/z-1}$ or through
the Kibble-Zurek mechanism~\cite{PhysRevB.99.094203}. In this respect, our results can be explored in state-of-the-art
experimental settings for moderate system sizes.
Our tentative approach draws a link between quantum information science and analog quasiperiodic systems without explicit order parameters, and it would be interesting to investigate whether our
results can be extended to more complex disordered models.

\begin{acknowledgements}
This work is supported by the National Natural Science Foundation of China (NSFC) under Grant No. 12174194, the startup fund of Nanjing University of Aeronautics and Astronautics under Grant No. 1008-YAH20006, Top-notch Academic Programs Project of Jiangsu Higher Education Institutions (TAPP), and stable supports for basic institute research under Grant No. 190101.
\end{acknowledgements}

\bibliography{references}

\end{document}